\documentclass{aa}
\usepackage{txfonts}
\usepackage{graphicx}
\begin{document}
   \title{The Globular Cluster System of \object{NGC\,4374}}

   \author{M. G\'omez
          \inst{1}
          \and
          T. Richtler
          \inst{2}
          }

   \offprints{M. G\'omez.}

   \institute{Depto. Astronom\'{\i}a y Astrof\'{\i}sica, P. Universidad Cat\'olica
                de Chile,
              Casilla 306, Santiago 22, Chile\\
              \email{mgomez@astro.puc.cl}
         \and
             Grupo de Astronom\'{\i}a, Departamento de F\'{\i}sica, 
                Universidad de Concepci\'on, Casilla 160-C,
                Concepci\'on, Chile\\
             \email{tom@coma.cfm.udec.cl}
             }

   \date{Received / Accepted}

   \abstract{We study the globular cluster system (GCS) of the giant elliptical
\object{NGC\,4374} (\object{M84}) in the Virgo cluster using B and R photometry. 
The colour distribution is bimodal with peaks at B-R=1.11 and B-R=1.36, fitting
well to those found in other early-type galaxies. 
The radial 
profile of the cluster number density is flatter than the galaxy light. Using the luminosity
function we derive a distance modulus of $\mu=31.61\pm0.2$, which within the uncertainty
agrees with the distance from surface brightness fluctuations. 
Blue and red clusters show similar radial concentrations and azimuthal distributions. 
The total number of clusters is  $N=1775\pm150$, which together with
our distance modulus leads to a specific frequency of $S_{N}=1.6\pm0.3$. This
value is surprisingly low for a giant elliptical, but resembles the case of 
merger remnants like \object{NGC\,1316}, where the low specific frequency is probably caused by the 
luminosity contribution of an intermediate-age population.
A further common property  is the high rate of type Ia supernovae
which also may indicate the existence of a younger population.
However, unlike in the case of \object{NGC\,1316}, one cannot find any further evidence 
that \object{NGC\,4374} indeed hosts younger populations. The low specific
frequency would also fit to a S0 galaxy seen face-on. 

\keywords{galaxies: distances and redshifts -- galaxies: elliptical and lenticular,
cD -- galaxies: individual: \object{NGC\,4374} -- galaxies: interactions -- 
galaxies: star clusters}
}

   \maketitle
%

\section{Introduction}

The study of globular clusters in early type galaxies
has reached a state, where surprises have become rare when
only individual galaxies are studied. Progress in understanding 
the relation between the morphology (i.e. the specific frequency of clusters, their
spatial as well as their
colour distribution) of a GCS
and the host galaxy properties normally emerges from analyzing larger
galaxy samples (e.g. Kundu \& Whitmore \cite{kundu01}; Larsen et~al. \cite{soren01}) 
However, from
time to time, one encounters galaxies which exhibit some
peculiarity in their GCS, which one would like to understand in a more general framework. 

For example, 
it is well known that ``normal'' elliptical galaxies have specific frequencies
higher than about 3 (see Sect.~\ref{sec:SpecFreq} 
for the definition and Elmegreen \cite{elmegreen99} for a review of
specific frequencies).
\object{NGC\,1316}, the brightest galaxy in the Fornax cluster, has nevertheless a
low specific frequency of only $\sim0.9$ (Grillmair et~al. \cite{grillmair99};
G\'omez et~al. \cite{gomez01}).
Since it is known as a merger remnant, it is tempting to seek an explanation
not in the low number of globular clusters, but in the high luminosity due
to the presence of an intermediate-age stellar population which formed in
the merger a few
Gyrs ago. Indeed, Goudfrooij et~al. (\cite{goud01a}), by means of spectroscopy 
of the brightest clusters, identified several
intermediate-age globular
clusters among them. Also the apparently high SN Ia rate (\object{NGC\,1316} 
hosted SN 1980N and SN 1981D)
could indicate 
a strong intermediate-age population, given that the progenitor population of
SNe Ia is suspected to have an age of a few Gyrs (Yungelson et~al. \cite{yungelson95}; 
Yoshii et~al. \cite{yoshii96}; McMillan \& Ciardullo \cite{mcmillan96}.)

The question arises whether the combination of a low specific frequency and
a high SN Ia rate is ubiquitous or whether more examples can be found pointing
to the possibility to use the specific frequency as an indicator for the presence
of a younger population.

Early-type galaxies, which hosted more than one SN Ia, are rare. Besides \object{NGC\,1316},
\object{NGC\,4374} in Virgo is one of the few ellipticals with two or even three SNe events:
1957B, 1980I (also labeled ``intergalactic'' due to its location
between \object{NGC\,4374} and \object{NGC\,4406}) and 1991bg.

The GCS of \object{NGC\,4374}  is not well investigated. Ajhar et~al. 
(\cite{ajhar94}) presented VRI photometry of globular clusters for
10 galaxies in Virgo and Leo ellipticals, among them \object{NGC\,4374}. 
Due to the small field, properties like the
specific frequency or the spatial distribution could not be addressed.  

In the light of a possible intermediate-age population, it is worthwhile to investigate
in detail the GCS of \object{NGC\,4374} with a larger field.

\section{Observations, reductions and photometry}

\subsection{Observations}
The observations were carried out at the 3.5m telescope at Calar Alto, Spain,
run by the Max-Planck Institute for Astronomy, Heidelberg.
The observation period was 19 to 21 March, 1999. The instrument was the
focal reducer MOSCA (www.caha.es/CAHA/Instruments/MOSCA/index.html) equipped
 with a 
Loral 2K~x~2K CCD. The pixel scale was 0.513 "/pix and the usable unvignetted field of view
  $\sim13'$~x~$13'$ (1.5K~x~1.5K). The filters in use were Johnson B and R.
In the first night several frames, centered on \object{NGC\,4374}, were acquired, as well 
as the Landolt fields SA98, SA101 and SA107 at different airmasses.
The observation log is given in Table~\ref{tab.observations}.

\begin{figure}
\centering
\includegraphics[width=8.8cm]{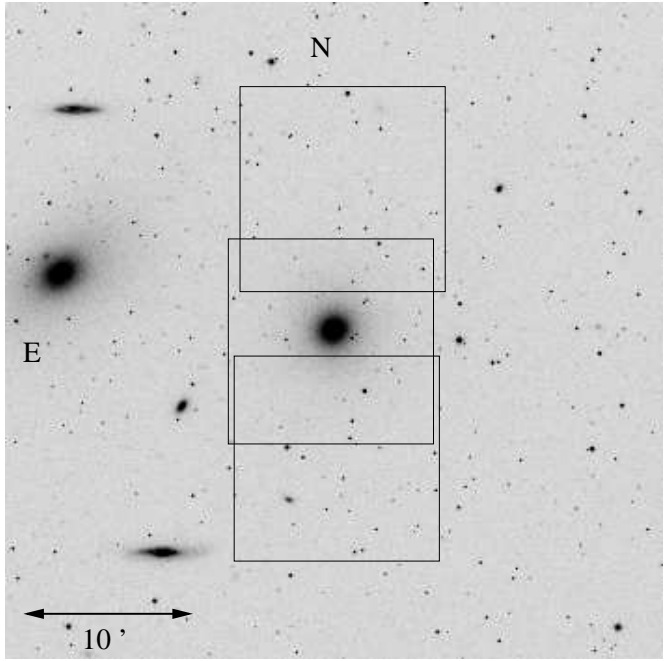}
      \caption{The observed fields in the three nights. \object{NGC\,4374} is at the center.
The bright elliptical at the left is NGC\,4406 (M86).}
       \label{fig:dss}
\end{figure}

\begin{table}
        \caption{The details of the observations. See Fig.~\ref{fig:dss} for the
                orientation of the fields.}
        \label{tab.observations}
        \begin{center}
        \begin{tabular}{llcc}
                \noalign{\smallskip}
                \hline
                \hline
                \noalign{\smallskip}
                night & field & B & R \\
                \noalign{\smallskip}
                \hline
                \noalign{\smallskip}
                19/3/99 & central & 30, 120, 5 x 600 s & 30, 120, 4 x 600 s\\
                20/3/99 & northern & 30, 120, 4 x 600 s & 30, 60, 4 x 600 s\\
                21/3/99 & southern & 120, 5 x 600 s & 60, 6 x 400 s\\
                \noalign{\smallskip}
                \hline
        \end{tabular}
        \end{center}

\end{table}

Two additional fields were also observed during the second and third night, with an
offset of $\sim8'$ to the north and south, respectively, as shown in Fig.~\ref{fig:dss}.

Bias subtraction and flat-fielding were performed using standard IRAF procedures, resulting in a  
flat-field accuracy of about $1\%$.
The images  were aligned using the centers of many bright stars as reference points. Bad pixels
were replaced by the average of 4 neighboring pixels. After this, we combined typically 5 
frames with exposure times 
ranging from 400 to 600 seconds in both B and R. Cosmics were removed by a $\sigma$-clipping algorithm.
The combined frames show a seeing of 1.7'' in B and 1.5'' in R.

A median filter was then applied to the combined frames to model and subtract
the galaxy light.

For the photometry, we used DAOPHOT under IRAF.
Daofind was run on each of these frames with a detection threshold of 3 times the $\sigma$
of the sky level. We then performed PSF photometry with {\it allstar}. The output lists were
matched to leave only objects detected in both filters.

For these objects we used the stellarity index of SExtractor 
(Bertin \& Arnouts \cite{bertin96}) for 
 distinguishing star-like objects from galaxies. 
The stellarity index ranges between 1.0 (star-like) and 
0.0 (extended). This classification, however, becomes
progressively more uncertain for fainter objects.

As we do not resolve globular clusters at the Virgo distance, 
 no light loss due to a different shape of clusters and stars
is observable. An aperture correction was applied in the usual way to calibrate our PSF-photometry
with that of the standard stars, which were measured using a much larger aperture.

\subsection{The calibration of the photometry}

Only the first of 3 nights was photometric. About 40 Landolt (\cite{landolt92})
standard stars were
observed at airmasses from 1.0 to 1.8, giving a total of $\sim 100$ datapoints.
Their colour range exceeds that of the globular clusters. We then performed 
aperture photometry with radii from 4 to 30 pixels. A curve-of-growth  was 
constructed for each standard, and only stars whose instrumental magnitudes converged
with increasing aperture radius were selected for the calibration.

The resulting transformation equations are:

\begin{eqnarray}
b = B + 0.874(\pm0.026) + 0.223(\pm0.021) \cdot X \nonumber\\
- 0.176(\pm0.021) \cdot(B-R) \\
r = R + 0.027(\pm0.022) + 0.096(\pm0.016) \cdot X \nonumber\\
+ 0.046(\pm0.020) \cdot(B-R)
\end{eqnarray}

\noindent
with $b,r,B,R$ the instrumental and standard magnitudes in the B and R filter.
$X$ is the airmass. The $rms$ of the
fit was 0.029 and 0.027 for $B$ and $R$, respectively.

The calibration has been compared with photoelectric aperture photometry of \object{NGC\,4374} 
published by
Poulain (\cite{poulain88}), using five aperture sizes. The mean difference was $\sim$0.02
mag for each band, without any systematic trend. Local standard stars were then defined
in the field to calibrate the non-photometric nights.

\begin{figure}
\centering
\includegraphics[width=8.7cm]{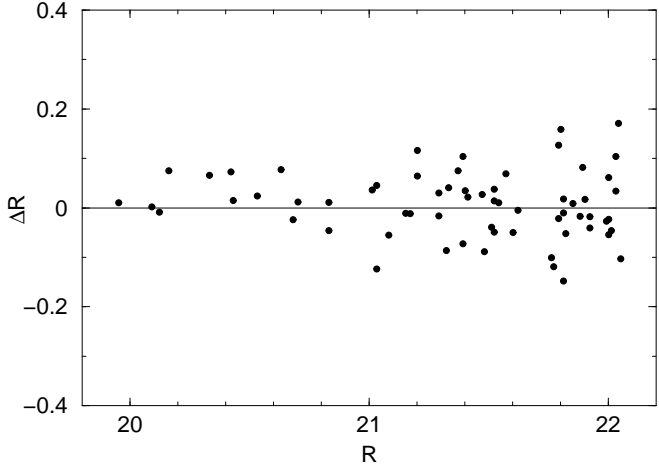}
      \caption{Comparison of common objects in the photometry of Ajhar et~al. (\cite{ajhar94}) and the present
               work in the R-band. $\Delta R$ means the magnitudes quoted by
               Ajhar et~al. minus ours. The mean difference is 0.006 mag with a $\sigma$ of 0.085 mag.} 
       \label{fig:AjharR}
\end{figure}

\begin{figure}
\centering
\includegraphics[width=8.8cm]{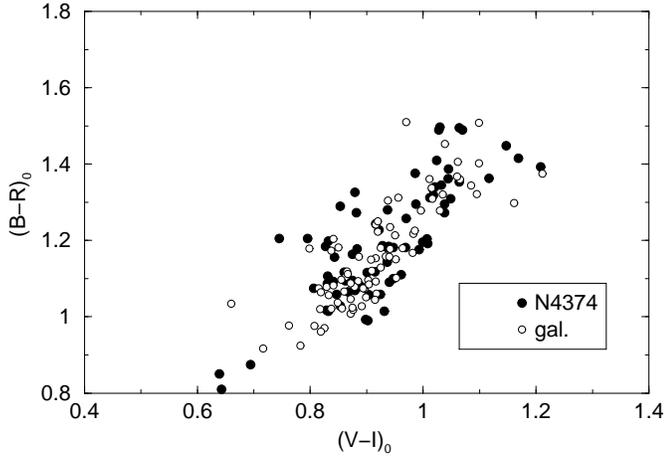}
      \caption{Colour comparison of common objects in the VRI photometry of Ajhar et~al. (\cite{ajhar94}) and our
       BR photometry (filled symbols). We chose the V-I colour because of its widespread use in the photometry of
        globular cluster systems. Overplotted are galactic globular clusters, where the dereddened colours are
      taken from the compilation of Harris (\cite{harris96}). There is no obvious offset between
      the clusters of \object{NGC\,4374} and galactic globular clusters, demonstrating  the quality of our B-R colours.}
       \label{fig:AjharVI}
\end{figure}

We also compared the R magnitudes of the
cluster candidates with about 100 clusters in common with  Ajhar et~al. (\cite{ajhar94}). The
mean difference is 0.006, with a $\sigma$ of 0.085. No trend with magnitude is
apparent (Fig.~\ref{fig:AjharR}). Regarding colours, no straightforward comparison is possible.
However, we can plot V-I colours of Ajhar et~al. versus our B-R colours and ask whether this relation
matches the (B-R)-(V-I) relation for galactic globular clusters. This is done
in Fig.~\ref{fig:AjharVI}.
The agreement is quite satisfactory. A few outliers may be caused by photometric errors. 

\subsection{Selection of cluster candidates}

886 objects have simultaneously  been detected  in the B and R bands. However, a fraction 
of them might be  foreground stars and background galaxies. 
To statistically select globular clusters among them, we have
used the following set of criteria:
\newcounter{marker}
\begin{list}{\roman{marker})}{\usecounter{marker}}
\item $B > 20.5$
\item $0.7 < B-R < 1.8$
\item uncertainty(B), uncertainty(B-R) $< 0.3$
\item stellarity index $> 0.35$
\end{list}

These criteria arise from the assumption that clusters around \object{NGC\,4374} 
should resemble the Galactic ones. Regarding the color range, we have used the
McMaster data (Harris \cite{harris96})
and chosen galactic
clusters with reddening smaller than $E_{B-V}=1.0$, resulting in the above
colour limits.

The colours were de-reddened and the magnitudes extinction-corrected 
using the maps of Schlegel et~al. (\cite{schlegel98}). We adopt 
$E_{B-V}=0.04$, $A_{B} = 0.173$ and $A_{R}=0.107$. In the following, only reddening 
corrected values appear.

\subsection{Completeness correction}

For the later determination of the  globular cluster luminosity function (GCLF)
we need to know what fraction of globular clusters is found in each
magnitude interval.
A common  way to evaluate the completeness is to perform artificial star 
experiments. Many artificial ``star-like''
objects are added to the frames and the detection, photometry and selection 
procedures are performed in exactly 
the same way as with the real cluster candidates.
The number of recovered and selected objects per magnitude bin, divided by the
initial number of artifical clusters 
gives an estimation of the completeness correction. 

To get sufficiently good
statistics, one needs a large number of artificial objects, but care must be 
taken not to increase the crowding. Therefore, consecutive tests with
a small number of artificial objects are preferred.

Using the known PSFs, we added 200 artificial clusters in steps of
0.1 mag, starting from B=20.0 down to B=26.0, and applied the same photometric
treatment and selection criteria as we did for 
the cluster candidates.

\begin{figure}
\centering
\includegraphics[width=8.8cm]{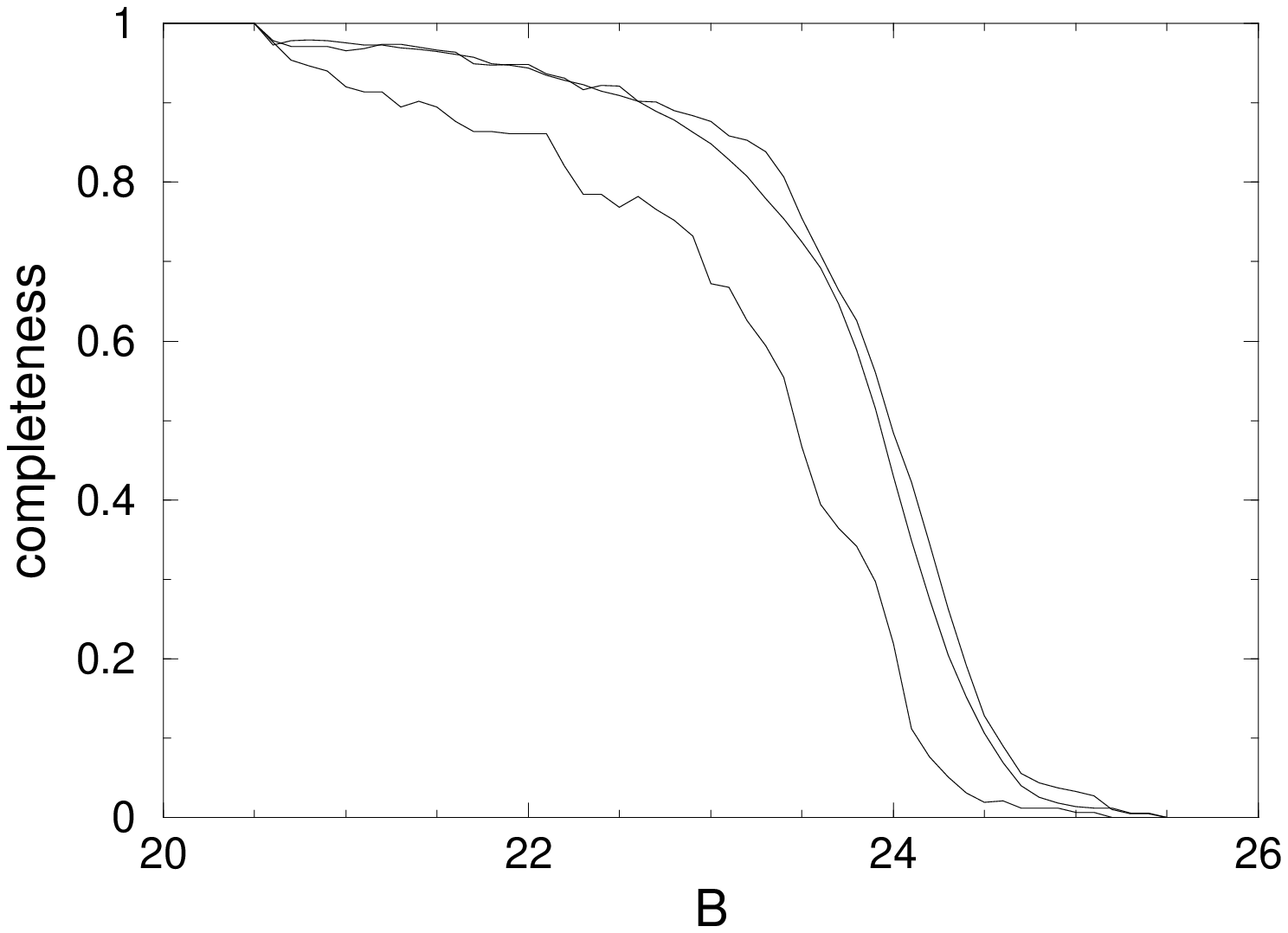}
\includegraphics[width=8.8cm]{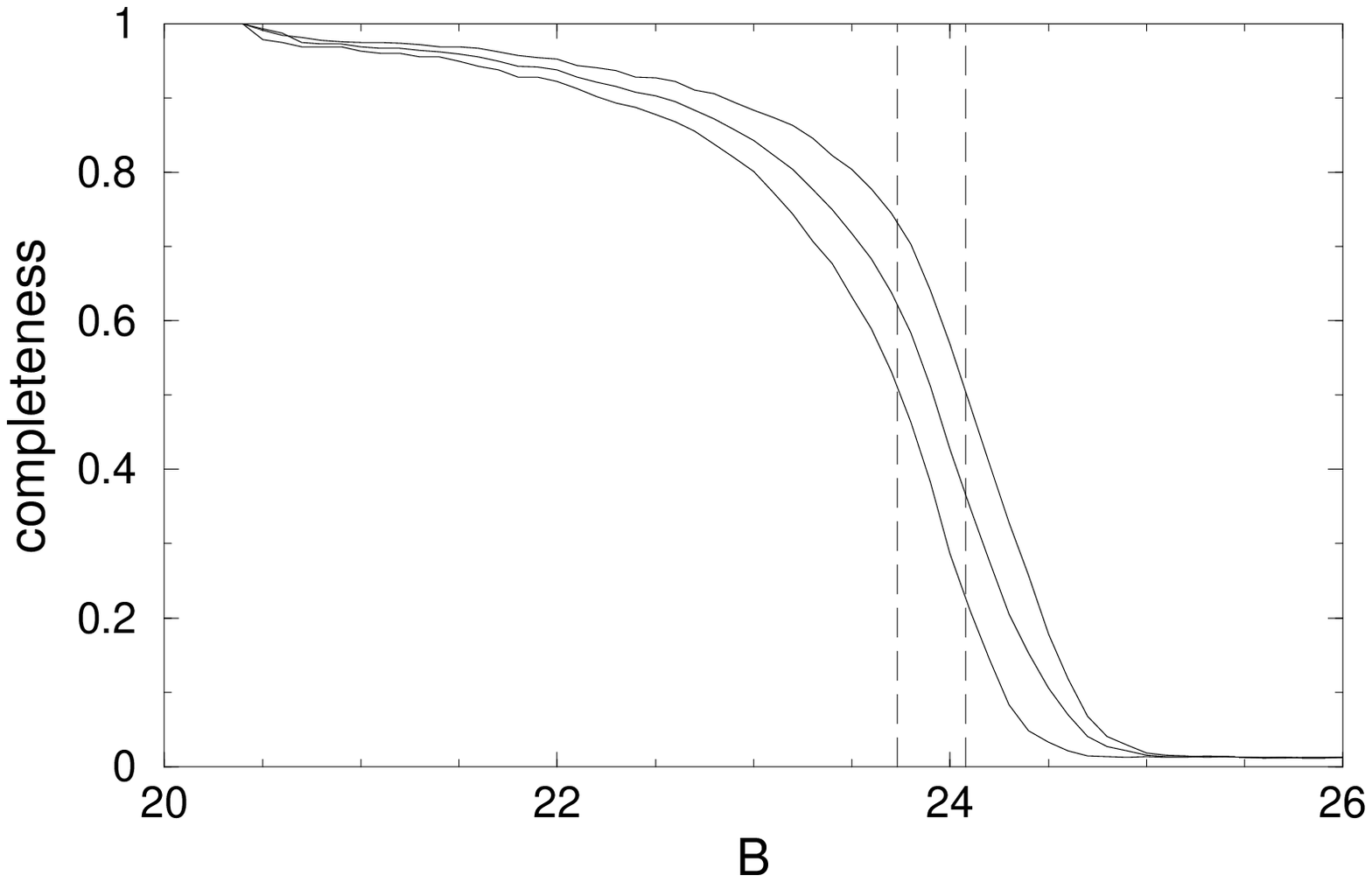}
      \caption{{\bf Top}: The completeness for three annuli centred 
        on \object{NGC\,4374}, as function
         of the B magnitude, for clusters having B-R=1.2. The annuli are described by 
        the following inner and outer radii 
        (in pixels): 50$<$r$<$150, 150$<$r$<$300 and r$>$300. Only the inner annulus has a 
        noticeable lower completeness. {\bf Bottom}: The mean completeness (averaged 
        over the three annuli), for clusters
        having colours of B-R=0.9, 1.2 and 1.5. These are represented by the three 
        solid lines from the left to the right. The shift with colour is evident. 
        The two dashed lines indicate the magnitude at which the mean completeness 
        falls to 0.5 for blue (left) and red (right) clusters. The difference 
        is $\sim$0.3 mag.} 
        \label{fig:voelligkeit}
\end{figure}

We expect the completeness to vary with galactocentric distance because of
the radially dependent galaxy light background. 
However, a decrease in the completeness was only noticed for clusters located between radii
of 50 and 150 pixels (25\farcs65 and 76\farcs95). At larger radii, the completeness corrections stays
constant with increasing distance.
Fig.~\ref{fig:voelligkeit} (upper panel) gives the resulting completeness in dependence
on B-magnitude for a colour of B-R = 1.2.

We estimated the dependence of the completeness on colour using clusters of
colours B-R=0.9, B-R=1.2 and B-R=1.5.
In all, 140000 clusters were added for each colour. Fig.~\ref{fig:voelligkeit} (lower panel)
shows that the completeness for red clusters is about 0.3 mag fainter than for blue
ones. This difference must be taken into account for the calculation of the radial 
profile of red and blue cluster candidates (Sect.~\ref{sec:blue_red}).

\subsection{Background correction}

Despite the applied selection criteria, there remain point sources,  
which may be foreground stars or 
unresolved background galaxies. The surface density profile levels out at a
radial distance of about 500\arcsec (see Fig.~\ref{fig:radialprof}). We therefore
have used the outermost parts of the two fields 
observed during the second and third nights (see Fig.~\ref{fig:dss}).
That is, two rectangular areas of 12\farcm8~x~4\farcm3 at the top of the upper field and at
the bottom of the lower field are our background fields.
The photometry for  these fields was done in the same way as for the central
field and it was calibrated using the local standards defined in the overlapping regions.

Completeness tests were also run on these two fields.
The two background populations were  averaged concerning colour distributions,
luminosity function and surface density.

\section{Results}

We first present the colour-magnitude diagram and the colour
distribution of those cluster candidates identified simultaneously in B and R.  
After that we study the morphological properties of blue and red candidates 
separately.
In these subsections we consider only sources brighter than B=23.8, where the
mean completeness is $\sim 60\%$ (see Fig.~\ref{fig:voelligkeit}).

For deriving the GCLF (Sect.~\ref{sec:gclf}) and the specific frequency 
($S_{N}$), we used only the R frame. 
The reason for doing so is that we reach
more than half a magnitude deeper using the R frames alone, which might be
crucial for the detection of the turn-over magnitude (hereafter TOM). 
We keep the criteria regarding magnitude range, photometric error and stellarity index.

Completeness and background corrections for the R frames have been performed
in the same way as for the combined sample.

\subsection{The colour-magnitude diagram}

\begin{figure}
\centering
\includegraphics[width=8.8cm]{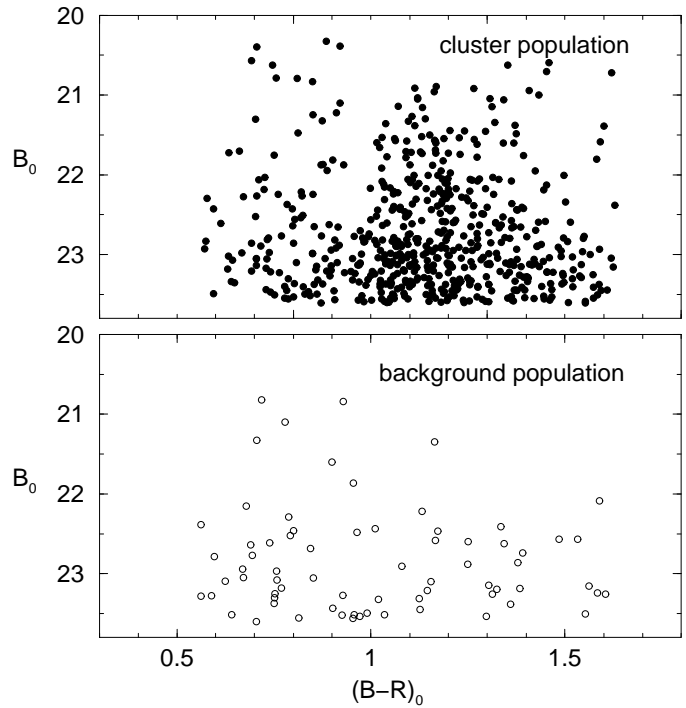}
      \caption{The B-R colour-magnitude diagram of cluster candidates around 
        \object{NGC\,4374} (top) and for the background population (bottom).}
       \label{fig:cmd}
\end{figure}

The B-R colour-magnitude diagram is shown  
in  Fig.~\ref{fig:cmd}. 
In the upper panel, cluster candidates brighter than
B=23.8 and having a radial distance less than 500\arcsec are plotted. 
The lower panel shows the same diagram for
the background population in an area which is smaller by a factor 1.3 than that for the
upper panel. It is apparent that objects bluer than B-R=0.9 are mainly foreground/background
objects. This is in good agreement with the fact that B-R=0.9 is also a limit
for the galactic clusters, as can be seen from Fig. \ref{fig:AjharVI}. 

\subsection{Colour distribution}

\begin{figure}
\centering
\includegraphics[width=8.8cm]{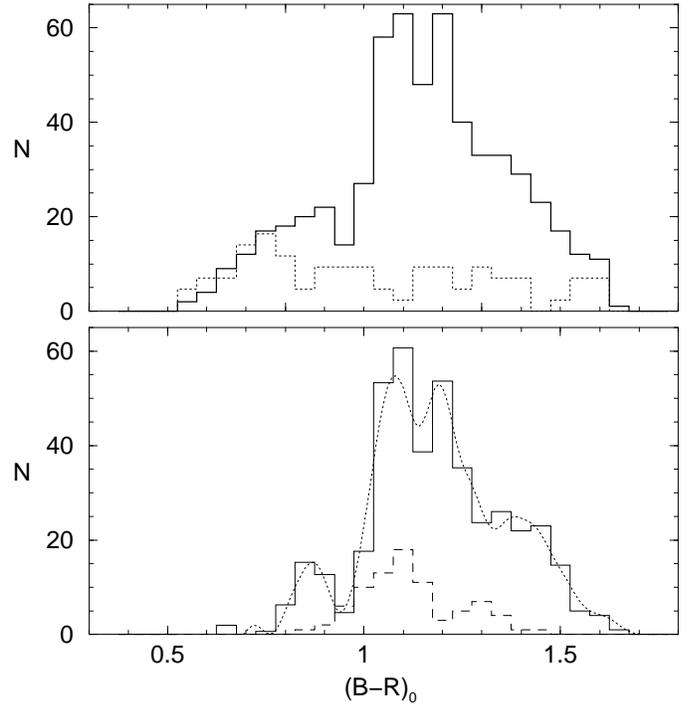}
      \caption{{\bf Top}: the B-R colour distribution of cluster candidates around 
        \object{NGC\,4374}. 
        The bin size is 0.05 mag. The thick solid line represents the colours
        before subtraction of the background population, whose distribution
	is shown with the dotted line.
        {\bf Bottom}: the B-R colour histogram after subtraction of the background population,
        together with a so-called ``generalised
        histogram'' (solid line). This is constructed by placing a gaussian in the
        abscissa at each
        B-R colour, and then adding the contribution from all gaussians together.
        The latter is independent of the bin center, and in our case agrees with the
         ``normal'' histogram. The dashed line shows the comparison with the galactic 
        clusters.}
       \label{fig:color}
\end{figure}

The colour histogram is given in Fig.~\ref{fig:color}. The upper panel shows
the colour distribution of all sources having a radial
distance less than 500\arcsec (bold line). 
The dotted line is the background normalised to the same area.
The lower panel shows the colour distribution for the
background corrected sample (solid line). For comparison the B-R distribution of
the galactic clusters is shown as well (dashed line), showing those clusters, for which B-R 
photometry is available and which have reddenings less than 1.0 (81 clusters
from the compilation of Harris \cite{harris96}).
The little peak at B-R=0.85 can be  a residual from an inappropriate
background subtraction. Without having further clues, we do not regard these
objects to be globular clusters.
 The galactic distribution has 
the peak at B-R=1.1, qualitatively agreeing with the blue peak of the
\object{NGC\,4374} clusters. 
 Although the comparison in the red regime is made difficult by the
fact that many metal-rich galactic clusters with higher reddenings are omitted,
it seems that the distribution of \object{NGC\,4374} clusters is more extended to the
red and hence presumably to more metal-rich clusters.
The appearance of the histogram does not depend much on the bin center  which
is demonstrated by
the dotted curve. This  is the sum of the
contributions from all clusters, where each one is represented by a Gaussian
at the corresponding colour. The only parameter is the $\sigma$ of the Gaussian
which we choose to be 0.05 to match the bin size.

There are not many B-R photometries of GCSs available, so a comparison only 
with 3 galaxies, \object{NGC\,1380} (Kissler-Patig et~al. \cite{kissler97}),
\object{NGC\,1199} 
and \object{NGC\,6868} (da Rocha et~al. \cite{darocha02}) is principally possible. 
Unfortunately, in all three
cases the colour distribution was not background corrected, which may hamper
the comparison. However, the blue peak
in these colour distribution was found at B-R=1.1 in all cases, resembling the
galactic system and, as we will see, that of NGC 4374.  

Ajhar et~al. (\cite{ajhar94}) and Gebhardt \& Kissler-Patig (\cite{gebhardt99})
report V-I photometry for GCs in \object{NGC\,4374}. Ajhar et~al. pointed out a strikingly narrow 
V-I colour distribution in comparison with other Virgo and Leo ellipticals, 
 but again a more detailed comparison is made difficult
by their small number statistics (moreover, no background subtraction has been
attempted). Gebhardt \& Kissler-Patig analyse the first 4 moments in the V-I 
colour distribution of a sample of 50 GCSs.  \object{NGC\,4374} shows one of
the highest skewness parameters in their sample, i.e. its colour distribution
is strongly weighted to the red.

The general appearance of the colour distribution looks bimodal. A bimodal 
colour distribution
has been found for many GCSs (e.g. Ashman \& Zepf \cite{ashmanzepf92};
Larsen et~al. \cite{soren01};
Kundu \& Whitmore \cite{kundu01}),
often interpreted as a signature of two episodes of cluster
formation, sometimes in the context of a merger scenario. However, see Dirsch
et~al. (\cite{dirsch03}) for the effects of a non-linear colour-metallicity relation.

\begin{figure}
\centering
\includegraphics[width=8.8cm]{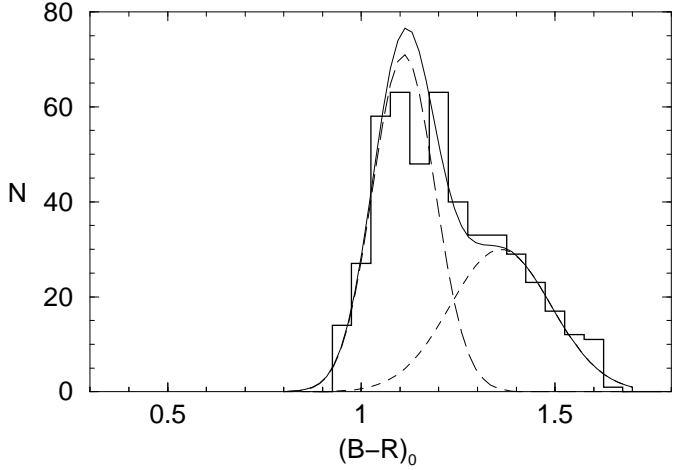}
      \caption{This plot shows the colour distribution of the GCS of \object{NGC\,4374}
       (not background corrected) together with the two Gaussians, which 
       best represent the distribution in the sense of a KMM-test. Only clusters
       redder than 0.925 have been included in the test. }
       \label{fig:KMM}
\end{figure}

Following a common statistical approach, we performed a KMM-test (Ashman,
Bird \& Zepf \cite{ashman94}) on the colour distribution. 
This test is applied to the distribution uncorrected for the background
population, in both homoscedastic and heteroscedastic versions. It returns 
that the colour distribution is best represented 
by two Gaussians with positions
at B-R=1.11 (the blue peak) and B-R=1.36 (the red peak), and 
with $\sigma$-values of 0.08 and 0.13, respectively (heteroscedastic mode). In the
homoscedastic case, the peaks are at 1.14 and 1.42 with $\sigma=0.10$.
The P-value, which gives the
probability of having a unimodal distribution, is practically zero in both versions, 
i.e. a unimodal distribution is excluded. Fig.~\ref{fig:KMM} shows the 
colour distribution together with the returned
Gaussians in the heteroscedastic fit.
In all GCSs investigated so far, the peaks exhibit approximately constant
colours. Weak dependence on galaxy luminosity has been found for the blue and
red peaks by Larsen et~al. (\cite{soren01}), while Forbes \& Forte (\cite{forbes01})
see only a dependence in case of the read peak.

As mentioned, also \object{NGC\,4374}
fits to this notion compared with GCSs, for which B-R photometry is available.
However, most work has been done in V-I (e.g. Larsen et~al. \cite{soren01};
Kundu \& Whitmore \cite{kundu01}) where the peaks are located at V-I=0.95 and V-I=1.15.     
As can be seen from Fig.~\ref{fig:AjharVI}, these values fit well to the B-R peak
colours quoted above.

\begin{figure}
\centering
\includegraphics[width=8.8cm]{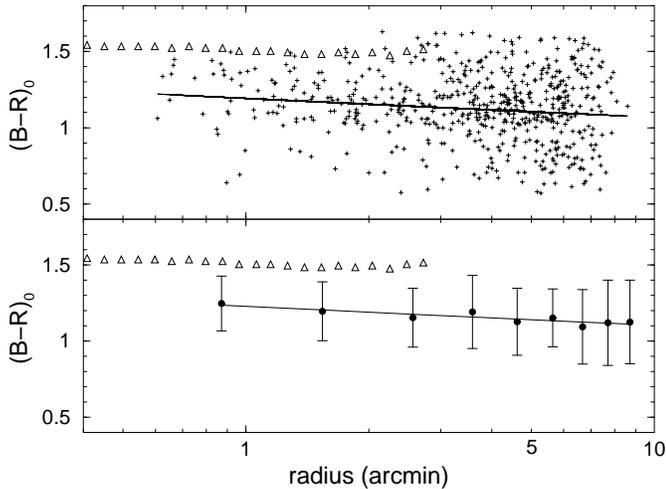}
      \caption{The colour gradient for cluster candidates brighter than B=23.8. The
upper panel shows the B-R colour of individual clusters vs. the projected distance to the optical
center. The B-R colour of the galaxy is shown for comparison (triangles). 
In the lower panel the mean color in concentric anuli (each 51\farcs3 wide) 
is plotted. The error bars are the $\sigma$ in the colour for each bin. 
The solid lines are the least-square fits.}
       \label{fig:colorgradient}
\end{figure}

Is there a colour gradient? Fig.~\ref{fig:colorgradient}  
plots 
the B-R colour versus the projected distance from the galaxy centre 
(in logarithmic scale).
The scatter is large, but a slight trend  that the clusters become
bluer at larger radii, may be recognizable.

\subsection{Blue and Red clusters}

\label{sec:blue_red}

We divided the clusters in two subpopulations at B-R=1.25 in order 
to search for possible morphological differences 
between the blue and the red clusters. 

\begin{figure}
\centering
\includegraphics[width=8.8cm]{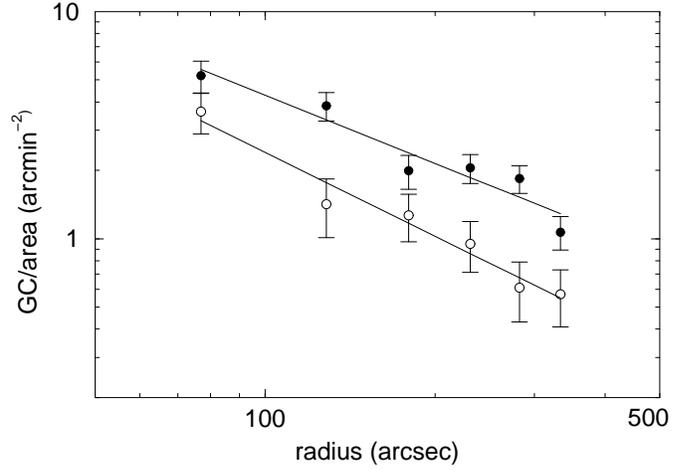}
      \caption{Radial profile of red (open circles) and blue clusters (filled circles), 
        for objects brighter than
         B=23.8. They are not clearly distinguishable in their concentration. The solid
        lines are least-square fits.}
       \label{fig:rp_blue_red}
\end{figure}

\begin{table}
        \caption{The counts for the radial profile of blue and red subsamples, using
        both B and R frames. The counts have been corrected for completeness and
        background contamination.}
        \label{tab.rp_blue_red}
        \begin{center}
        \begin{tabular}{lccc}
        \noalign{\smallskip}
        \hline
        \hline
        \noalign{\smallskip}
        r [\arcsec] & area [$\sq\arcmin$] & $N^\mathrm{blue}$ & $N^\mathrm{red}$ \\
        \noalign{\smallskip}
        \hline
        \noalign{\smallskip}
        77\farcs0  & 6\farcm890  & 35.9 $\pm$ 5.8 & 25.0 $\pm$ 5.1        \\
        128\farcs3 & 11\farcm483 & 44.2 $\pm$ 6.3 & 16.3 $\pm$ 4.7        \\
        179\farcs6 & 16\farcm076 & 32.0 $\pm$ 5.4 & 20.4 $\pm$ 4.9        \\
        230\farcs9 & 20\farcm669 & 42.4 $\pm$ 6.3 & 19.7 $\pm$ 4.9        \\
        282\farcs2 & 25\farcm262 & 46.4 $\pm$ 6.5 & 15.4 $\pm$ 4.6        \\
        333\farcs5 & 29\farcm856 & 32.1 $\pm$ 5.4 & 17.1 $\pm$ 4.8        \\
        \noalign{\smallskip}
        \hline
        \end{tabular}
        \end{center}
\end{table}

The resulting counts are given in Table~\ref{tab.rp_blue_red}, 
and the radial profiles of red and blue clusters
are shown in Fig.~\ref{fig:rp_blue_red} after the subtraction of the corresponding red
and blue background levels. Red and blue clusters do not
significantly differ in their radial distributions.
They are well represented
by: $\rho(r)_\mathrm{red} \propto r^{-1.22 \pm 0.12}$ and 
$\rho(r)_\mathrm{blue} \propto r^{-1.00 \pm 0.15}$.

\begin{figure}
\centering
\includegraphics[width=8.8cm]{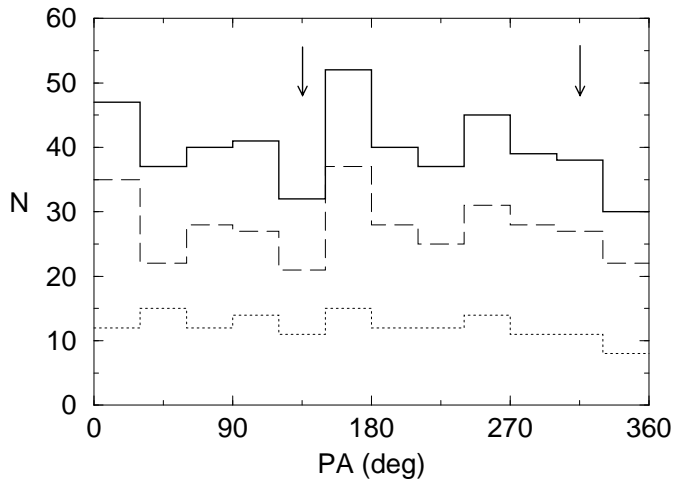}
      \caption{The azimuthal distribution of the clusters. The position angle is 
counted north over east. The histogram at the top (solid
line) represents the number of cluster candidates on each sector. The binning angle is
30 degrees. The dashed and dotted lines are the angular distribution of blue and red 
clusters, respectively. No systematic difference exists and both red and blue clusters
show a circular distribution. The galaxy's position angle of the major axis is indicated by
the arrows at 135 and 315 degrees.}
       \label{fig:angular}
\end{figure}

To investigate possible differences regarding the azimuthal distribution, we counted the
cluster candidates in
12 equally-sized sectors, each spanning an angle of 30 degrees. The region inside 
the radius r=51\farcs3 (or,
equivalently, $\sim5$ kpc) was left out as the completeness is significantly
lower. Clusters outside $r=350\arcsec$ were not considered because the sectors become
geometrically incomplete.
The results (Fig.~\ref{fig:angular}) show that, on average, red and blue clusters are spherically
distributed around \object{NGC\,4374}. The galaxy's position angle of the major axis is indicated by
the arrows at 135 and 315 degrees.   

There are some cases where blue and red clusters show significant differences in their
azimuthal distributions, for example \object{NGC\,1316} (G\'omez \& Richtler \cite{gomez01}), 
\object{NGC\,3115} (Kavelaars \cite{kavelaars98}), and \object{NGC\,1380} (Kissler-Patig et~al.
\cite{kissler97}). In these galaxies, the red clusters resemble the shape of their 
host galaxies, while the blue clusters are more spherical. The interpretation as
differences between halo and bulge/disk populations is suggestive. 
The elongations of the above galaxies, if caused by inclined disks, are strong
enough to make a disk population of clusters easily distinguishable from a
spherical halo population. This is not the case for \object{NGC\,4374} with
its low ellipticity. Clearly, we cannot draw conclusions
about halo and bulge/disk cluster subpopulations from their azimuthal
distribution.

\subsection{Radial Profile}

We calculated the cluster radial profile by counting
the cluster candidates in several annuli centered on \object{NGC\,4374}, each 25\farcs65 wide, 
starting from a radius of $r=$89\farcs8 up to $r=$346\farcs4, and 102\farcs6 wide thereafter.
The innermost ring ($r=$77\farcs0) was given a width of 51\farcs3 because of the low
statistics.
The results are listed in Table~\ref{tab.rp}. The bin center and size
are the first two columns. The area of the annuli is given in the third column.
The fourth column shows the number of candidates up to the 60\% detection limit,
before correcting for the completeness. The corrected number of clusters and its
density are given in columns five and six. The remaining columns are used in the
derivation of the specific frequency $S_{N}$ (see Sect.~\ref{sec:SpecFreq}).

\begin{table*}
  \caption{This table shows the result of the radial profile for the surface density
    of cluster candidates. The first three columns list the
    center of each annulus (in arcsecs), its size and its area, in square arcmin. 
    The observed number of
    cluster candidates brighter than R=23.5 on the corresponding annuli is given
    in column~4. After correcting for completeness, the actual number is shown in 
    column~5. Column~6 lists this number divided by the area of the annulus,
    or the surface density of clusters. The next three columns are used for the
    derivation of the specific frequency $S_{N}$. They are the number of candidates 
    up to the observed TOM before and after the completeness 
    correction (columns 7
    and 8), and the total number of clusters in each annulus, assuming the LF to
    be symmetric around the TOM (column~9). The background level has been subtracted
    for these latter counts. For radii larger than 333\farcs5, 
    the annuli contain only background sources, so these four last rows are
    left blank.}
  \label{tab.rp}
  \begin{center}
    \begin{tabular}{lccrccrccc}
      \noalign{\smallskip}
      \hline
      \hline
      \noalign{\smallskip}
      r [\arcsec] & size [\arcsec] & area [$\sq\arcmin$] & $N^\mathrm{(R)}_\mathrm{obs}$ & $N^\mathrm{(R)}_\mathrm{corr}$ 
      & GC$/$area [$1/\sq\arcmin$] & $N^\mathrm{(R)}_\mathrm{TOM}$ & $N^\mathrm{(R)}_\mathrm{TOM,corr}$
      & $N_\mathrm{ring}$  \\
      \noalign{\smallskip}
      \hline
      \noalign{\smallskip}
      77\farcs0  & 51\farcs3   & 7\farcm022  & 167  & 182 & $25.92\pm1.91$   & 168 & 183 & $249\pm14$ \\
      102\farcs6 & 25\farcs7   & 4\farcm683  & 95   & 104 & $22.21\pm2.18$   & 95  & 104 & $130\pm11$ \\
      128\farcs3 & 25\farcs7   & 5\farcm850  & 106  & 116 & $19.83\pm1.84$   & 106 & 118 & $138\pm11$ \\
      153\farcs9 & 25\farcs7   & 7\farcm027  & 105  & 115 & $16.37\pm1.52$   & 106 & 117 & $117\pm11$ \\
      179\farcs6 & 25\farcs7   & 8\farcm198  & 127  & 139 & $16.96\pm1.43$   & 127 & 140 & $143\pm12$ \\
      205\farcs2 & 25\farcs7   & 9\farcm367  & 139  & 152 & $16.23\pm1.32$   & 139 & 153 & $149\pm13$ \\
      230\farcs9 & 25\farcs7   & 10\farcm534 & 152  & 166 & $15.76\pm1.22$   & 152 & 167 & $158\pm13$ \\
      256\farcs5 & 25\farcs7   & 11\farcm706 & 147  & 160 & $13.67\pm1.08$   & 148 & 162 & $128\pm14$ \\
      282\farcs2 & 25\farcs7   & 12\farcm876 & 144  & 157 & $12.19\pm0.97$   & 144 & 158 & $101\pm13$ \\
      307\farcs8 & 25\farcs7   & 14\farcm050 & 153  & 167 & $11.89\pm0.92$   & 155 & 170 & $105\pm13$ \\
      333\farcs5 & 25\farcs7   & 15\farcm222 & 147  & 160 & $10.51\pm0.83$   & 147 & 161 & $67\pm14$ \\
      397\farcs6 & 102\farcs6  & 50\farcm378 & 458  & 480 & $9.52\pm0.43$    & --- & --- & --- \\
      500\farcs2 & 102\farcs6  & 29\farcm210 & 241  & 257 & $8.80\pm0.55$    & --- & --- & --- \\
      602\farcs8 & 102\farcs6  & 24\farcm332 & 201  & 211 & $8.67\pm0.57$    & --- & --- & --- \\
      705\farcs4 & 102\farcs6  & 23\farcm701 & 199  & 209 & $8.82\pm0.61$    & --- & --- & --- \\

      \noalign{\smallskip}
      \hline
    \end{tabular}
  \end{center}
\end{table*}

\begin{figure}
\centering
\includegraphics[width=8.8cm]{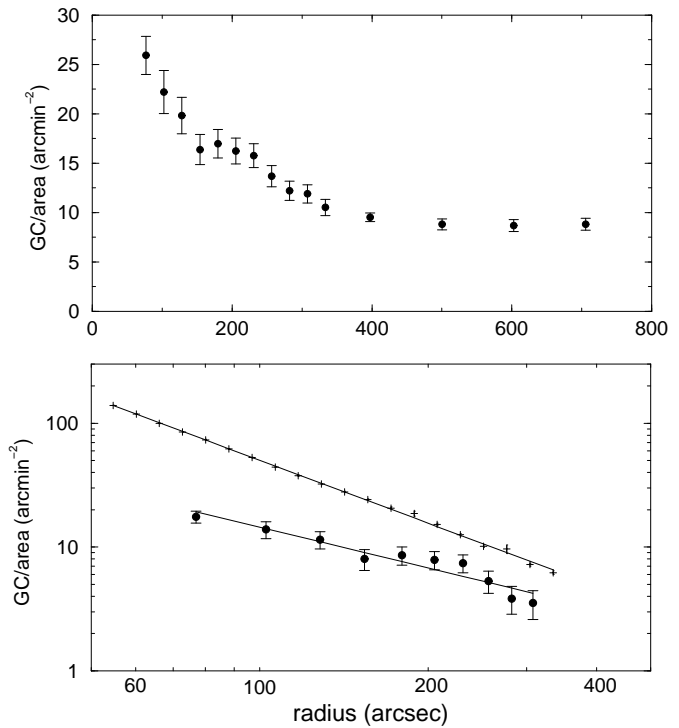}
      \caption{The radial profile of the clusters' surface density for candidates brighter
than R=23.5. {\bf Top}: the number of
candidates per square arcmin vs. the mean projected distance to the optical center.
{\bf Bottom}: the radial profile, after the correction for background objects. The galaxy light,
arbitrarily shifted (crosses) and the least-squares fits (solid lines) are also plotted.} 
       \label{fig:radialprof}
\end{figure}

The radial profile in R is shown in Fig.~\ref{fig:radialprof}. The upper panel
plots the GC number surface density and it is apparent that the cluster 
population extends at least out to $r=350$''.
At larger radii the background dominates the profile.

The background level is well determined and its value of $8.36\pm0.60$ arcmin$^{-2}$
has been subtracted from the tabulated densities, to derive the  
true
radial profile, which is shown in the lower panel. The galaxy light in the R-band,
arbitrarily shifted, is also plotted. The solid lines are:
$\rho(r)_\mathrm{gcs} \propto r^{-1.09 \pm 0.12}$
and $\rho(r)_\mathrm{gal} \propto r^{-1.67 \pm 0.02}$ 
, with $\rho(r)_\mathrm{gcs}$ being the cluster surface density (in units of arcmin$^{-2}$)
and $\rho(r)_\mathrm{gal}$ the  surface brightness of the galaxy light.

It is also apparent that the radial profile of the cluster density around \object{NGC\,4374} is
flatter than the galaxy light.

\subsection{The Luminosity Function}

\label{sec:gclf}

The GCLF has been widely used as a distance indicator
(see Harris \cite{harris01} and Richtler \cite{richtler03} for recent reviews). 
GCLFs of a broad range of galaxy types are
well characterised by log-normal or $t_{5}$ functions, the latter being of the form:

\begin{equation}
t_{5}(m) = \frac{8}{3\sqrt{5}\pi\sigma_{t}}
  \left(1+\frac{(m-m_0)^{2}}{5\sigma_{t}^{2}} \right)^{-3}
\end{equation}

Ideally, the photometry should reach  deeper than
 $m_{0}$, the turnover magnitude (TOM).
Unfortunately, our combined photometry in B and R turns out to be insufficient
to reach the expected TOM for the Virgo cluster. The results of our artificial stars 
experiments show that we go
half a magnitude fainter using only R instead of both B and R. For this reason, we
restrict to the R-band in the derivation of the GCLF.

\begin{figure}
\centering
\includegraphics[width=8.8cm]{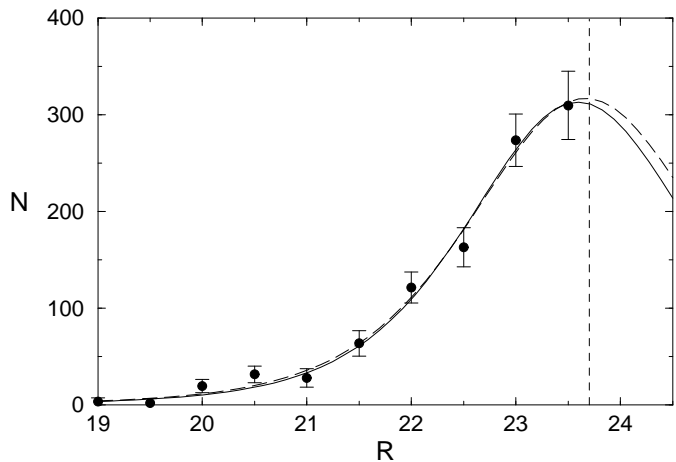}
      \caption{The GCLF in the R band. The solid line is the least-square fit of a $t_{5}$ 
        function (equation 3) keeping $\sigma=1.1$. The derived TOM is $23.56\pm0.15$. If
        $\sigma$ is left a free parameter, the resulting TOM is $23.84\pm0.35$ and the
        function is given by the dashed curve. 
        The vertical line at R$\sim23.7$ indicates the magnitude at which the average 
        completeness is 50\% using only R-band measurements.}
       \label{fig:gclf}
\end{figure}

The counts are listed in Table~\ref{tab.gclf}. We remark that a gaussian fit does not give systematically
different results (see e.g. Della Valle et~al. \cite{dellavalle98}; Larsen et~al. \cite{soren01}; 
Kundu \& Whitmore \cite{kundu01}), but a $t_{5}$ function
leads to a better representation of our data. Furthermore, we assume a $\sigma_{t}$ of 1.1
which is a typical value for bright ellipticals (Harris \cite{harris01}; Larsen et~al. 
\cite{soren01}). 
The least-square fit is shown
in Fig.~\ref{fig:gclf}, where the data are the number of cluster candidates (corrected 
for completeness and background contamination) at each 0.5 magnitude in R. The error bars account
for poisson error in the binning process and for the completeness correction. According
to this fit, the TOM in the R band is 23.56$\pm0.15$.

\begin{table*}
\begin{center}
        \caption{The counts for the GCLF in the R band. The first column is the bin
        center. The second column gives the observed number of candidates. The
completeness factors (in column~3) lead to the completeness-corrected number of
candidates in column~4. The observed background population and background completeness 
are listed in columns 6 and 7. The completeness correction for the background field is slightly
smaller than for the central field due to better seeing. Finally, after subtraction
of the background population (scaled to the area of the clusters), the total number of
clusters is given in column~8.}
        \label{tab.gclf}
        \begin{tabular}{lccccccc}
        \noalign{\smallskip}
        \hline
        \hline
        \noalign{\smallskip}
        R & $N_\mathrm{obs}$ & $f_\mathrm{field}$ & $N_\mathrm{corr}$ & 
        $N^\mathrm{bkg}_\mathrm{obs}$ & $f_\mathrm{bkg}$ & $N^\mathrm{bkg}_\mathrm{corr}$
        & $N_\mathrm{t}$ \\
        \noalign{\smallskip}
        \hline
        \noalign{\smallskip}
        19.0  & 12 &  0.980  & 12.3 $\pm 3.5$   & 8   & 0.995 & 8.5 $\pm 2.0$    & 0.9  $\pm 4.0$\\
        19.5  & 20 &  0.975  & 20.5 $\pm 4.6$   & 12  & 0.995 & 12.2 $\pm 2.5$   & 4.2  $\pm 5.2$\\
        20.0  & 39 &  0.975  & 40.0 $\pm 6.4$   & 17  & 0.990 & 17.4 $\pm 2.9$   & 16.6 $\pm 7.0$\\
        20.5  & 63 &  0.965  & 65.3 $\pm 8.2$   & 27  & 0.985 & 27.3 $\pm 3.7$   & 28.7 $\pm 9.0$\\
        21.0  & 79 &  0.955  & 82.7 $\pm 9.3$   & 39  & 0.965 & 40.2 $\pm 4.6$   & 28.8 $\pm 10.4$\\
        21.5  & 150 & 0.945  & 158.7 $\pm 13.0$ & 68  & 0.965 & 70.3 $\pm 6.0$   & 64.5 $\pm 14.3$\\
        22.0  & 204 & 0.920  & 221.7 $\pm 15.5$ & 80  & 0.945 & 84.9 $\pm 6.7$   & 108.0 $\pm 16.9$\\
        22.5  & 290 & 0.885  & 327.7 $\pm 19.2$ & 123 & 0.940 & 130.7 $\pm 8.3$  & 152.5 $\pm 20.9$\\
        23.0  & 423 & 0.785  & 538.9 $\pm 26.2$ & 181 & 0.905 & 200.1 $\pm 10.5$ & 270.7 $\pm 28.2$\\
        23.5  & 440 & 0.675  & 651.9 $\pm 31.1$ & 211 & 0.790 & 266.6 $\pm 13.0$ & 294.6 $\pm 33.7$\\
        24.0  & 278 & 0.330  & 842.4 $\pm 50.5$ & 157 & 0.465 & 338.4 $\pm 19.0$ & -- \\
        24.5  & 121 & 0.155  & 780.6 $\pm 71.0$ & 66  & 0.175 & 378.5 $\pm 32.8$ & -- \\

        \noalign{\smallskip}
        \hline
        \end{tabular}
\end{center}
\end{table*}

Leaving $\sigma_t$ a free parameter, we get for the TOM(R) = 23.84$\pm$0.35 and $\sigma_t$ = 1.22$\pm$0.11.
We take the absolute TOM(R) from Della Valle et~al. (\cite{dellavalle98}) who quote $M_R = -8.14 \pm 0.07$ for the
galactic system. Another absolute calibration comes from the Andromeda nebula. Barmby et~al. (\cite{barmby01})
give R=16.40$\pm0.14$ for the apparent TOM of \object{M31}. Together with a distance modulus of $\mu=24.38\pm0.07$
(Freedman et~al. \cite{freedman01}), one gets $M_R = -7.98\pm0.15$. We average these two values and adopt
$M_R = -8.06\pm0.15$.

This leads to distance moduli of $31.61\pm0.20$ (keeping $\sigma_t=1.1$) and $31.90\pm0.38$
(leaving $\sigma_t$ free) for \object{NGC\,4374}. In the following, we consider only the TOM with $\sigma_t=1.1$.

The SBF distance modulus is $\mu=31.32\pm0.11$ (Tonry et~al. \cite{tonry01}).
Our distance, although marginally higher, is still in agreement within the uncertainties. A slightly higher
distance would anyway be expected because we used zeropoints from the Galaxy and \object{M31}, which contain
a larger fraction of blue clusters. Larsen et al. (\cite{soren01}) find that the V-TOMs of blue clusters
are brighter than those of red clusters by an  average of 0.37 mag in their sample of 17 early-type galaxies. 
This difference should be somewhat smaller in R. 
 Our photometry
does not reveal the TOM definitely. However, it only could be fainter. If this were the case the difference
between the SBF distance and the TOM distance would grow larger. There are some examples where the disagreement
between SBF and GCLF distance can be related to the existence of an intermediate-age population 
(Richtler \cite{richtler03}). Therefore, a deeper photometry would be of interest.

\subsection{Specific Frequency}

\label{sec:SpecFreq}

The specific frequency relates the number of clusters belonging to a given
galaxy with its luminosity in units of $M_{V}=-15$. It was first
introduced by Harris \& van~den~Bergh (\cite{harris81}) as:

\begin{equation}
S_N = N\cdot10^{0.4\cdot(M_V+15)}
\end{equation}

\noindent
with $N$ the total number of clusters and $M_{V}$ the absolute
magnitude of the galaxy in the $V$ band.

Once the turnover has been measured, $M_{V}$ is normally derived via a calibration
with the Galactic GCS. Neither $M_{V}$ nor $N$ are
trivial quantities to measure. As it is nearly impossible to observe the total 
population of clusters,
$N$ must be determined through rough extrapolations. Typically, the number of candidates
brighter than the TOM is obtained (or estimated, if the TOM is not reached by
the observations), and then the counts are doubled, assuming the LF to be
represented by a symmetric function (a $t_{5}$ in our case).
Matters are further complicated if the observations do not reach the background level,
in which case, a spatial extrapolation is required as well.

The radial profile indicates
that our population of candidates extends out to a projected radius of $\sim6$ 
arcmin. This region
is fully covered by our frames, so we do not need any extrapolation down to the
background level.

Again, we have used only the R frame to count the clusters in several annuli,
in almost the same way as with the radial profile. The only difference is that
we count now up to the TOM (R=23.56).
Table~\ref{tab.rp} lists the counts for $S_{N}$. The bin center and size
are given in the first two columns. The area of the annulus (in arcmin$^{2}$) is
in the third column.
The seventh column shows the observed number of candidates up to the TOM. 
The completeness-corrected number of clusters in each annulus is in column~8.
Finally, assuming the LF to be symmetric around the TOM and after the correction
for background objects (8.36 arcmin$^{-2}$), the total number of candidates on 
each annulus is given in column~9.

We have left out the inner region of the galaxy ($r<$51\farcs3), 
because the completeness decreases dramatically as a consequence of the high noise 
associated to the galaxy light. To estimate the contribution of this region to the
total number of clusters, we have used the power-law describing the radial profile.
This gives 65 candidates per arcmin$^{2}$, or 166 candidates in this region
down to the TOM.

Adding the contribution of all annuli gives $N=1775\pm150$ candidates distributed
over a total area of 112.53 arcmin$^{2}$.
The distance modulus derived in the previous section is $\mu=31.61 \pm 0.2$ and the 
total apparent magnitude of \object{NGC\,4374} in the $V$ band is 9.11 (de~Vaucouleurs et~al. 
\cite{devaucouleurs91}). Using the colour
excess of $E_{B-V}=0.04$ (Schlegel et~al. \cite{schlegel98}), the absolute magnitude
is $M_{V}=-22.62 \pm 0.2$; we therefore derive a specific frequency of $S_{N}=1.6\pm0.3$. 
This is a very low value for an early-type galaxy of this luminosity class. 
A possible explanation is suggested in the discussion.

\section{Discussion}

\object{NGC\,4374} is similar to \object{NGC\,1316} in two aspects which could
both indicate 
the presence of an intermediate age-population: a high rate of supernova (SN) type Ia and
a low specific frequency of globular clusters. SN type Ia are believed to
have their progenitors in populations older than 1 Gyr (e.g. Yungelson et~al. 
\cite{yungelson95}; Yoshii et~al. \cite{yoshii96}; McMillan \& Ciardullo
\cite{mcmillan96})  and
a low $S_{N}$ value of 1.6 might result from the higher galaxy luminosity produced by an
intermediate-age population with respect to older populations. 

In the case of \object{NGC\,1316}, this has been already proven by the identification
of GCs with ages of about 3 Gyr (Goudfrooij et al. \cite{goud01a}). These clusters probably formed in a major merger
event which was accompanied by a starburst. The very high star formation rate
during these mergers seems to provide suitable conditions for the formation
of massive stellar clusters (Larsen \& Richtler \cite{soren00}).     

For \object{NGC\,4374}, spectra of clusters are not yet available, and broad-band photometry remains
inconclusive regarding the existence of intermediate-age populations.

Integrated spectroscopy as well gives no indication: 
Terlevich \& Forbes (\cite{terlevich02})
have catalogued the ages and metallicities
of $\sim$150 field and cluster galaxies. Using the SSP models from Worthey \& Ottaviani 
(\cite{worthey97})
and the
H$\beta$ index measured by Goudfrooij et~al. (\cite{goud99}), they quote an age of 11.8 Gyr
for \object{NGC\,4374}. The quoted ages for merger remnants like \object{NGC\,1316}
and \object{NGC\,5018} (host to the Ia supernova 2002dj) 
 are 3.4 Gyr and 1.5 Gyr, respectively. \object{NGC\,5018} also shares with \object{NGC\,4374}
 the characteristic of a low specific frequency (Hilker \& Kissler \cite{hilker96}).  

However, \object{NGC\,4374} is a polar-ring elliptical (Bettoni et~al. \cite{bettoni01};
note also that among the 10 polar-ring ellipticals, listed in that work,
3 are Ia host galaxies).
It has a dust lane, which extends out to about 
1 kpc, and is a prominent radio source as well
(see Hansen et~al. \cite{hansen85} for details on the morphology of the
dust lane). These properties might be related to previous merger processes.

An interesting note regarding Ia host galaxies and GCSs
can be made from the paper of Gebhardt \& Kissler-Patig
(\cite{gebhardt99}). These authors analyse the V-I colour distribution of the GCs
of a sample of early-type galaxies. Their ``skewness'' parameter measures the
asymmetry of the colour distribution with respect to the mean colour. The two
GCSs which are skewed strongest towards red (i.e. metal-rich) clusters both
belong to Ia host galaxies (\object{NGC\,4536}, \object{NGC\,4374}) as well as does the fourth in this
sequence (\object{NGC\,3115}).

Since SNe Ia in ``normal'' elliptical galaxies are so rare, one might ask
whether \object{NGC\,4374} is in fact a S0 galaxy seen face on. This would alleviate the
problem of the low specific frequency, and given the uncertainties,
a specific frequency of 1.9 (the upper limit) is not unusual
for a S0 galaxy (Harris \cite{harris91}; Kundu \& Whitmore \cite{kundu01}).

Looking for more peculiarities, it is striking that the ellipticity of the isophotes
of \object{NGC\,4374} decreases with increasing radial distance rather than the opposite, which is
commonly observed for ellipticals (Caon et~al. \cite{caon94}). It is plausible that
a face-on disk, which in projection in the inner parts is dominated by an
elliptical bulge, may show this feature.

In that case, one may expect more conclusive information from dynamical analyses.  
Gerhard et~al. (\cite{gerhard01}) 
included \object{NGC\,4374} in their sample of 21 round early-type galaxies for which they have
derived the mass profiles out to 1 to 1.5 effective radii.  According to
Magorrian \& Ballantyne (\cite{magorrian01}), a flattened structure when seen face-on, but
dynamically treated as spherical, shows a spurious radial bias in its orbit
structure. \object{NGC\,4374}, however, does not show any striking anomaly regarding its
anisotropy parameters. On the other hand, the radial extension of the
analysis is limited to 100 arcsec of radial distance, so data which reach
further out are of high interest. Investigating the kinematics of globular clusters is
the best way  to pursue this suspicion.

\section{Conclusions}

We analysed the GCS of the giant elliptical \object{NGC\,4374}, host of two or even
three SNe Ia. This may lead to the suspicion that an intermediate-age population
may be present in \object{NGC\,4374}, and perhaps manifest in the GCS properties.  
However, the colour distribution does not show anomalous features. 
It is best
represented by a bimodal distribution  with peaks at B-R=1.11 and B-R=1.36, resembling the
colour distribution found in other early-type galaxies.  A unimodal
distribution is  ruled out by a KMM test.

The  spatial distributions  of the red and blue 
subpopulations do not show significant differences. 

However, the number of clusters in \object{NGC\,4374} turns out to be surprisingly
lower than what is expected for its high luminosity. This could indicate  the presence of
an intermediate-age population, as  has been the case for other early-type
galaxies with low specific frequencies. It is intriguing that these examples
(\object{NGC\,1316} and \object{NGC\,5018})  also have been hosts of type Ia SNe.
 
The other possibility is that \object{NGC\,4374} is a S0 galaxy seen face-on, in which case
the low specific frequency of 1.6 would fit to other objects of this class. One then would 
expect a disk-like kinematic behaviour of the  metal-rich globular clusters or
of a subsample . Measuring 
radial velocities of many globular clusters  
could provide further clues.

\begin{acknowledgements}

We thank an anonymous referee for his/her comments which considerably improved the 
paper, particularly for pointing out an error in the photometric calibration.
 We also thank Georg Drenkhahn for discussions.
We have made use of the NED Database, as well as online-data for \object{NGC\,4374} hosted 
by CDS. T.R. acknowledges support from the FONDAP center for astrophysics, Conicyt 15010003. 
M.G. thanks the DAAD for a studentship.
\end{acknowledgements}

\end{document}